\documentclass[twocolumn,showpacs,preprintnumbers,amsmath,amssymb]{revtex4}

\usepackage{graphicx}
\usepackage{dcolumn}
\usepackage{bm}

\begin{document}

\preprint{PRL/BPC}

\title{Stages of Homogeneous Nucleation in Solid Isotopic Helium Mixtures}

\author{M. Poole}
\author{J. Saunders}
\author{B. Cowan}

\email{b.cowan@rhul.ac.uk}

\affiliation{Millikelvin Laboratory, Royal Holloway University of
London, Egham, TW20 0EX, UK.}

\date{\today}

\begin{abstract}
We have made pressure and NMR measurements during the evolution of
phase separation in solid helium isotopic mixtures. Our observations
indicate clearly all three stages of the homogeneous nucleation --
growth process: 1) creation of nucleation sites; 2) growth of the
new-phase component at these nucleation sites; and 3) coarsening:
the dissolution of sub-critical droplets with the consequent further
late-stage growth of the super-critical droplets. The time exponent
for the coarsening, $a = 1/3$, is consistent with the conserved
order parameter Lifshitz-Slezov evaporation-condensation mechanism.
\end{abstract}

\pacs{67.80.Jd  68.35.Rh}
\maketitle

Nucleation and phase separation in binary mixtures are central to
the understanding of phenomena ranging from the strength of metallic
alloys\cite{McKie1974} to the properties of
polymers\cite{Cabral2002} and they provide examples of first-order
phase transitions with a rich parameter space. While second-order
transitions are relatively well-understood, with universality
classes for both static and dynamical critical
behaviour\cite{Hohenberg1977}, the situation is not so clear in the
first-order case. The nature of the order parameter is of vital
importance; in particular the kinetics will depend crucially on
whether the order parameter is is conserved or
not\cite{Mazenko1985}. It has been proposed that scaling concepts
may be as useful in the study of first-order transitions as they
have proved to be in the second-order case\cite{Mazenko1985}.
Universality in late-stage `coarsening' and evidence for scaling of
the structure factor have been demonstrated
theoretically\cite{Katano1984}, but other questions remain
open\cite{Onuki2002}.

Phase separation in solid helium mixtures is an example of a
first-order transition with a conserved order parameter (COP). This
system is attractive because the segregation process occurs on an
accessible time scale: slower that that in fluids, but faster than
that in conventional solids. This is a consequence of the unique
nature of the atomic motion in solid helium where quantum exchange
results in a temperature-independent diffusion coefficient,
intermediate between that of a solid and a
liquid\cite{Bennemann1976}.


In a first-order phase transition, fluctuations provide the energy
to surmount the barrier separating the initial and evolving phases.
This is the nucleation-growth scenario of Cahn and
Hilliard\cite{Cahn1958}, extended by
others\cite{Lifshitz1959,Slezov1997}. The barrier height is
determined by a balance between surface and volume energies, and
this gives a critical droplet size. Embryos larger than this
critical size are stable and grow. This leads to the identification
of three distinct stages of the nucleation-growth
process\cite{Slezov1997}: 1) creation of nucleation sites or
embryos; 2) growth of the new-phase component at these sites at the
expense of the background matrix; and 3)`coarsening' of the
super-critical droplets at the expense of the dissolving
sub-critical droplets when the background matrix is strongly
depleted. In our system the timescales are such that all three
stages may be identified distinctly.

Most experimental research on phase separation has been performed on
metallic alloys and in polymer mixtures. The Al-Zn alloy has been
studied by Mainville \emph{et al}.\cite{Mainville1997} using
small-angle x-ray scattering on critical mixtures. And the surface
of polymer mixtures has been studied by the Higgins group using
light scattering, neutron scattering and atomic force
microscopy\cite{Cabral2002}. Late-stage coarsening of crystallizing
biological macromolecules has been observed by Ng \emph{et
al}.\cite{Ng1996}, with a time scale of three to four months.

Previous work\cite{Penzev2002,Smith2003} has demonstrated the
initial stages of phase separation in high quality crystals of solid
helium mixtures. In this Letter we report the first observation of
all three stages of homogeneous nucleation and growth in this COP
system. The three stages of the nucleation-growth process have been
observed in the magnetic ordering of Cu$_3$Au
alloy\cite{Nagler1988}. However that is a non-conserved order
parameter (NCOP) system with, correspondingly, different kinetics.


Our experimental approach involves the use of two powerful tools:
NMR and high precision pressure measurements, both utilized during
step-wise cooling through the transition allowing the observation of
phase separation in real time. $^{3}$He NMR exploits the dependence
of the spin-lattice and spin-spin relaxation times $T_1$ and $T_2$
on $^{3}$He concentration\cite{Greenberg1972}. The relaxation times
depend on the strength of the inter-nuclear dipolar fields and on
their time modulation\cite{Cowan1997}.

The signature of two evolving phases, due to phase separation, is
the appearance of two components in measurements of both $T_1$ and
$T_2$. Each relaxation profile is well-described by a double
exponential form, allowing the proportions of the two evolving
phases to be inferred. Further, in the presence of a magnetic field
gradient, measurements of bounded diffusion within the concentrated
phase allows determination of both the spin diffusion coefficient
and the size of the droplets during the separation process. The
minimum droplet size that may be determined in this way scales with
$(D/\gamma G)^{1/3}$, the numerical coefficient depending upon
instrumentation details. In our experiments, for the liquid droplets
discussed here $l_{\rm{min}} \sim1 \mu \rm{m}$, while for solid
droplets with a smaller diffusion coefficient $l_{\rm{min}} \sim 0.1
\mu \rm{m}$\cite{Smith2003}.

Phase separation at constant volume results in a pressure increase,
a consequence of the excess volume\cite{Mullin1968} associated with
nonlinearity of the mixture molar volume as a function of
composition. The sample pressure is measured by a capacitative
strain gauge, providing in principle a continuous record of the
phase separation following a cooling step and a determination of the
characteristic time constant of the process.

Studies of homogeneous nucleation were performed on 1\%, 2\%, 7\%
and 50\% $^{3}$He mixtures. All stages of homogeneous nucleation
were investigated at 2\% and we concentrate on these results here.
At this concentration there there is a large accessible meta-stable
region of the phase diagram, allowing an appreciable temperature
quench without entering the spinodal region\cite{Edwards1989}. (By
contrast there is no meta-stable region at 50\% concentration; there
we have observed phase separation by spinodal
decomposition\cite{Poole2006}). The crystal was grown at a pressure
of 28 bar, corresponding to a molar volume of 20.9~cm$^3$. At this
pressure the temperature at which phase separation starts was found
to be 295 mK and new-phase regions form as $^{3}$He-rich liquid
droplets. In order to ensure good crystallographic quality the
crystal was grown at constant pressure; x-ray measurements have
shown this method to produce crystals of high
quality\cite{Fraass1987}. The procedure, together with other
experimental details have been described previously\cite{Smith2003}.
We studied this crystal in two separate cooling cycles. In the first
we lowered the temperature from above the transition in small steps
of 5$-$10mK; in the second we quenched in one step to 100mK.

Homogeneous nucleation in a uniform supersaturated mixture proceeds
through the formation of clusters of the new phase at random sites.
If the number of particles $n$ in a cluster is smaller than the
critical value $n_{\mathrm{c}}$ then it is unstable and it decays.
When $n>n_{\mathrm{c}}$, however, it grows. For a spherical cluster
$n_{\mathrm{c}}$ is given by\cite{Smith2003,Slezov1997}
\begin{equation}
n_{\mathrm{c}}=\left(  \frac{\beta}{\ln\left(
c_{0}/c_{\mathrm{f}}\right) }\right)^{3}
\end{equation}
where $c_{0}$ is the initial $^{3}$He concentration of the mixture,
$c_{\mathrm{f}}$ is the concentration at the final temperature
$T_{\mathrm{f}}$ and
\begin{equation}
\beta=\frac{8\pi}{3}\frac{\sigma
a^{2}}{k_{\mathrm{B}}T_{\mathrm{f}}}.
\end{equation}
Here $\sigma$ is the surface tension at the cluster boundary, $a$ is
the inter-atomic distance and $k_{\mathrm{B}}$ is Boltzmann's
constant. In our first step through the transition $n_{\mathrm{c}}$
will be about 40, corresponding to $ l \sim 1\mathrm{nm}$, too small
to observe by NMR methods.

The nucleation rate is a very rapid function of the degree of
supersaturation. As a consequence the formation of new nuclei occurs
only in the early part of the phase separation process. This results
in a maximal cluster concentration (per site) during a cooling step
given by\cite{Slezov1997}
\begin{equation}
N_{\mathrm{m}}=\sqrt{2}c_{0}^{7/4}\left( \frac{3}{2\pi\beta}\right)
^{3/8}\exp\left(  -\frac{3\beta^{3}}{8\ln^{2}\left(  c_{0}/c_{\mathrm{f}%
}\right)  }\right)  .
\end{equation}
The subsequent growth of new-phase droplets occurs by diffusion of
$^{3}$He atoms, through the background matrix, to the droplets. The
characteristic time for this growth process is
\begin{equation}
\tau_{D}=\frac{a^{2}}{3D}c_{0}^{-1/3}N_{\mathrm{m}}^{-2/3}%
\end{equation}
where $D$ is the diffusion coefficient of the $^{3}$He in the
matrix.

\begin{figure}
\includegraphics[width = 2.7 in]{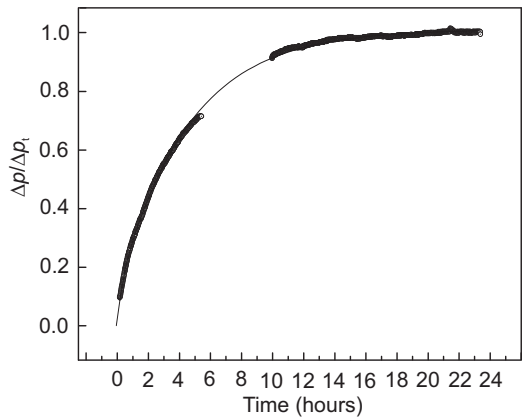}
\caption{\label{fig:fig_1} Time dependence of the sample pressure
following a step lowering of temperature from 220.2~mK to 210.8~mK.
The gap in experimental points is associated with a helium transfer
where the pressure gauge becomes unreliable.}
\end{figure}

In the step-wise cooling experiment we lowered the temperature from
above the transition in 5$-$10 mK steps: 295.0 mK, 290.0 mK, 285.3
mK, 280.4 mK, 275.8 mK, 270.2 mK,... down to 150.0 mK. We monitored
the pressure increase following each step; an example is shown in
Fig.~1. The line through the points is an exponential relaxation
fit, from which the time constant $\tau_{D}$ for evolution may be
found. From knowledge of $\tau_{D}$, using Eq.~(4), $N_{\mathrm{m}}$
the density of nuclei produced during the step may be found. Here
$c_0$ is determined from the phase diagram profile\cite{Edwards1989}
and the diffusion coefficient $D$ of the $^3$He in the $^4$He
background is known from Grigor'ev \emph{et
al.}\cite{Grigorev1997,Ganshin1999}. These calculations are similar
to those in our work on solid droplets\cite{Smith2003}; further
details are provided there.

The treatment is based upon the assumption that the total volume of
the new-phase droplets is sufficiently small that it has negligible
effect on the degree of supersatuation. This allows a `linear'
approach to the kinetics of the process where the growth of a
droplet is independent of the state of the others.

We found that the majority of droplets were nucleated in steps~2
and~3; in the second step there were $0.78 \times 10^{-15}$ droplets
per site produced, while in the third step there were $2.18 \times
10^{-15}$ produced. In each of the later steps there were less than
$10^{-17}$ produced. In this way we have identified the first stage
of the separation process during these earlier steps.

During the latter steps the number of $^3$He atoms in the
concentrated phase is observed by NMR to increase, as
$c_{\mathrm{f}}$ decreases in accordance with the phase diagram
profile. However since we have established that there are negligible
new droplets nucleating, we conclude that the existing droplets
continue to grow. Thus we identified the second stage of the
nucleation-growth process.

Once $N_{\mathrm{m}}$ is known one may determine the droplet surface
tension at each temperature by solving Eq.~(3) for $\beta$ and then
using Eq.~(2) to find $\sigma$. The result is that the surface
tension of the liquid droplets is found to be of order $10^{-5}$ J
m$^{-2}$; values for different temperatures are shown in Fig.~2.
These values are consistent with the inter-phase surface tension
measurements in phase-separated bulk liquid helium solutions of
Ohishi \emph{et al.}\cite{Ohishi1998} (at lower pressures), also
shown in the figure. At low temperatures the data tend towards the
$T^2$ dependence, in accordance with Fermi Liquid theory. The jump
in the points at 255~mK corresponds to a structural transition in
the background matrix\cite{Edwards1989}, from bcc at higher
temperatures to hcp at lower temperatures. The larger value of
$\sigma$ is consistent with the greater number of surrounding atoms
in the denser hcp phase.

An independent measure of droplet concentration is provided by the
droplet size inferred by NMR. At low temperatures, when all the
$^3$He resides in the droplets, the droplet diameter was determined
to be $14.0 \pm 1.5~\mu$m. Then for a 2\% initial $^3$He
concentration this would indicate a total droplet concentration of
$5.5 \pm 1.8 \times 10^{-15}$. Now the total droplet concentration
determined by pressure measurements is the sum of those from each
step, giving $3.0 \times 10^{-15}$. This evaluation supports the
inference that nucleation occurs in the first few steps. Precise
numerical agreement is precluded by uncertainties about the shape of
droplets. The inferred spin-diffusion coefficient of the $^3$He
liquid droplets and its temperature dependence are in good agreement
with those of the bulk liquid.
\begin{figure}
\includegraphics[width = 2.5 in]{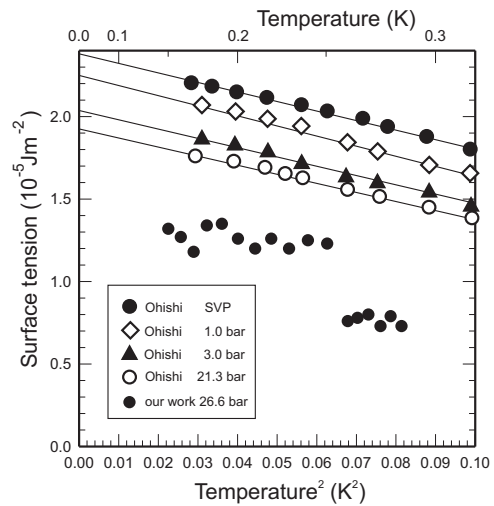}
\caption{\label{fig:fig_1} Surface tension of the droplets as a
function of temperature. Also show is the surface tension of the
interface in bulk liquid mixtures measured by Ohishi at lower
pressures.}
\end{figure}

The size of the droplets depends on the depth of the cooling step
through the transition. A small step gives a small degree of
super-saturation -- a lower nucleation rate and thus fewer nuclei.
However the number of atoms in the new phase is determined, from the
phase diagram, by the temperature. Thus a smaller cooling step will
result in larger droplets. Droplet size also depends on crystal
quality\footnote{The observation of homogeneous nucleation requires
high crystal quality. Crystalline imperfections favor heterogeneous
nucleation, resulting in more and therefore smaller droplets. Koster
\emph{et al.}\cite{Koster1995} find a droplet size of 25nm
comparable with the crystallite size determined from x-ray
scattering.}.


The linear approximation to the droplet growth, above, is not valid
in the final stages of the separation process where the degree of
supersaturation becomes small. In that case the critical droplet
size becomes large. This has two consequences. Firstly there is no
nucleation of new droplets: the probability of an adequate
fluctuation for this to happen becomes vanishingly small. Secondly,
as the critical droplet size becomes larger there is an increasing
number of droplets that find themselves smaller than this critical
size. They {\it were} stable; they are now unstable. These
sub-critical droplets will dissolve, liberating ${^3}$He atoms into
the background. And these atoms are then available to condense on
the remaining super-critical droplets they encounter.

During this late-stage growth the droplet size is predicted to
increase with a characteristic power law $l(t) \sim
t^{a}$\cite{Mazenko1985}, where the exponent $a$ depends on the
`universality class' of the transition. To investigate this
`coarsening' we quenched the crystal from above the separation down
to 100~mK and followed the droplet size using NMR. The evolution of
$l$ with time is shown in Fig.~3 where we have plotted $1/l$ against
$1/t^{1/3}$. The asymptotic late-stage behavior is indicated at the
left hand side where the approach to linearity indicates the
characteristic exponent $a = 1/3$, in accordance with the
Lifshitz-Slezov law\cite{Lifshitz1959} for a COP process.

This asymptotic behavior follows from simple scaling
arguments\cite{Jones2002}. The droplet curvature $\sim l(t)^{-1}$
will lead to concentration gradients of magnitude $\sigma/l(t)^2$.
We assume that the transport of ${^3}$He atoms arises through
diffusion, of coefficient $D$. Then the diffusive flux is $\sim D
\sigma/l(t)^2$. And it is this flux of atoms that results in the
final growth of the super-critical droplets, so that
${\rm{d}}l(t)/{\rm{d}}t \sim D \sigma/l(t)^2$, which has solution
$l(t) \sim {(\sigma D t)}^{1/3}$. A full solution of the actual
equations of motion\cite{Lifshitz1959} leads to the expression
\begin{equation}
l(t) = \left( {\frac{4}{9}\frac{{vc_\infty  }}{{k_{\rm{B}} T}}\sigma
Dt} \right)^{1/3}
\end{equation}
where $v$ is the atomic volume and $c_\infty$ is the equilibrium
concentration of ${^3}$He in the dilute phase.

\begin{figure}
\centerline{\includegraphics[width=2.5in]{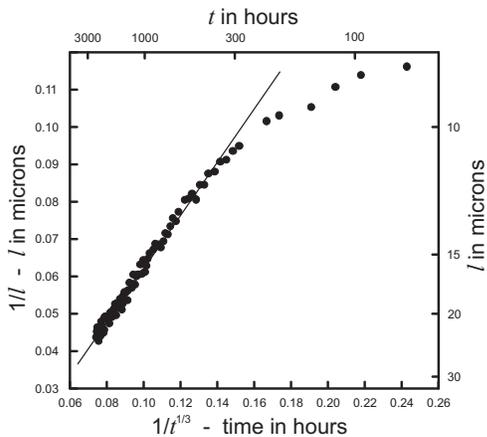}}
\caption{\label{fig:Ost} Late-stage coarsening.}
\end{figure}

The slope of the asymptote of Fig.~3 is found to be $1.09 \times
10^{7}$ s$^{1/3}$ m$^{-1}$. This may be compared with the value
calculated from Eq.~(5). The Edwards and Balibar phase diagram
equations\cite{Edwards1989} give $c_\infty = 4.5 \times 10^{-6}$ at
100mK. At these low concentrations $D$ is given\cite{Ganshin1999}
approximately by $D_0/c_\infty$, where $D_0 \sim 9.2 \times
10^{-12}$ m$^2$ s$^{-1}$. And this leads to an asymptote slope of
$({9k_{\rm{B}} T}/{4v c_\infty \sigma D})^{1/3}=0.89 \times 10^7
\;{\rm{s}}^{1/3} {\rm{m}}^{ - 1}$, within 20\% of the experimental
value. This is a reasonable agreement since the value of $D_0$ is
approximate; there are uncertainties in its magnitude and pressure
dependence\cite{Ganshin1999}. Thus we have observed the final,
coarsening, stage of the phase separation process.

In summary we have presented a detailed real-time investigation of
homogeneous nucleation in solid helium mixtures, in which all three
stages of the process have been identified. The time exponent for
the coarsening stage, $a = 1/3$ is in accord with the
Lifshitz-Slezov COP evaporation-condensation mechanism in contrast
to the $a=1/2$ behavior observed by Nagler \emph{et
al.}\cite{Nagler1988} for their NCOP transition. These observations
lend support to the proposal of universal exponents for late-stage
coarsening. Solid helium mixtures thus provide a model system for
the study of nucleation and growth kinetics in first-order phase
transitions.

Future work will include further investigation of early-stage
kinetics and a determination of the critical droplet size using
small-angle neutron scattering in high quality mixture crystals.

\begin{acknowledgments}
The authors wish to thank J\'{a}n Ny\'{e}ki and Tom Crane for
experimental assistance. This work was supported by EPSRC grant
EP/E023177/1.
\end{acknowledgments}

\bibliography{homogeneous}

\begin{thebibliography}{26}
\expandafter\ifx\csname natexlab\endcsname\relax\def\natexlab#1{#1}\fi
\expandafter\ifx\csname bibnamefont\endcsname\relax
  \def\bibnamefont#1{#1}\fi
\expandafter\ifx\csname bibfnamefont\endcsname\relax
  \def\bibfnamefont#1{#1}\fi
\expandafter\ifx\csname citenamefont\endcsname\relax
  \def\citenamefont#1{#1}\fi
\expandafter\ifx\csname url\endcsname\relax
  \def\url#1{\texttt{#1}}\fi
\expandafter\ifx\csname urlprefix\endcsname\relax\def\urlprefix{URL }\fi
\providecommand{\bibinfo}[2]{#2}
\providecommand{\eprint}[2][]{\url{#2}}

\bibitem[{\citenamefont{McKie and McKie}(1974)}]{McKie1974}
\bibinfo{author}{\bibfnamefont{D.}~\bibnamefont{McKie}} \bibnamefont{and}
  \bibinfo{author}{\bibfnamefont{C.}~\bibnamefont{McKie}},
  \emph{\bibinfo{title}{Crystalline Solids}} (\bibinfo{publisher}{Nelson,
  London}, \bibinfo{year}{1974}).

\bibitem[{\citenamefont{Cabral et~al.}(2002)\citenamefont{Cabral, Higgins,
  Yerina, and Magonov}}]{Cabral2002}
\bibinfo{author}{\bibfnamefont{J.~T.} \bibnamefont{Cabral}},
  \bibinfo{author}{\bibfnamefont{J.~S.} \bibnamefont{Higgins}},
  \bibinfo{author}{\bibfnamefont{N.~A.} \bibnamefont{Yerina}},
  \bibnamefont{and} \bibinfo{author}{\bibfnamefont{S.~N.~E.}
  \bibnamefont{Magonov}}, \bibinfo{journal}{Macromolecules}
  \textbf{\bibinfo{volume}{35}}, \bibinfo{pages}{1941} (\bibinfo{year}{2002}).

\bibitem[{\citenamefont{Hohenberg and Halperin}(1977)}]{Hohenberg1977}
\bibinfo{author}{\bibfnamefont{P.~C.} \bibnamefont{Hohenberg}}
  \bibnamefont{and} \bibinfo{author}{\bibfnamefont{B.~I.}
  \bibnamefont{Halperin}}, \bibinfo{journal}{Rev. Mod. Phys.}
  \textbf{\bibinfo{volume}{49}}, \bibinfo{pages}{435} (\bibinfo{year}{1977}).

\bibitem[{\citenamefont{Mazenko et~al.}(1985)\citenamefont{Mazenko, Valls, and
  Zhang}}]{Mazenko1985}
\bibinfo{author}{\bibfnamefont{G.~F.} \bibnamefont{Mazenko}},
  \bibinfo{author}{\bibfnamefont{O.~T.} \bibnamefont{Valls}}, \bibnamefont{and}
  \bibinfo{author}{\bibfnamefont{F.~C.} \bibnamefont{Zhang}},
  \bibinfo{journal}{Phys. Rev. B} \textbf{\bibinfo{volume}{31}},
  \bibinfo{pages}{4453} (\bibinfo{year}{1985}).

\bibitem[{\citenamefont{Katano and Iizumi}(1984)}]{Katano1984}
\bibinfo{author}{\bibfnamefont{S.}~\bibnamefont{Katano}} \bibnamefont{and}
  \bibinfo{author}{\bibfnamefont{M.}~\bibnamefont{Iizumi}},
  \bibinfo{journal}{Phys. Rev. Lett.} \textbf{\bibinfo{volume}{52}},
  \bibinfo{pages}{835} (\bibinfo{year}{1984}).

\bibitem[{\citenamefont{Onuki}(2002)}]{Onuki2002}
\bibinfo{author}{\bibfnamefont{A.}~\bibnamefont{Onuki}},
  \emph{\bibinfo{title}{Phase Transition Dynamics}}
  (\bibinfo{publisher}{Cambridge University Press}, \bibinfo{year}{2002}).

\bibitem[{\citenamefont{Bennemann and Ketterson}(1976)}]{Bennemann1976}
\bibinfo{author}{\bibfnamefont{K.~H.} \bibnamefont{Bennemann}}
  \bibnamefont{and} \bibinfo{author}{\bibfnamefont{J.~B.}
  \bibnamefont{Ketterson}}, \emph{\bibinfo{title}{The Physics of Liquid and
  Solid Helium}} (\bibinfo{publisher}{Wiley, New York}, \bibinfo{year}{1976}).

\bibitem[{\citenamefont{Cahn and Hilliard}(1958)}]{Cahn1958}
\bibinfo{author}{\bibfnamefont{J.~W.} \bibnamefont{Cahn}} \bibnamefont{and}
  \bibinfo{author}{\bibfnamefont{J.~E.} \bibnamefont{Hilliard}},
  \bibinfo{journal}{J. Chem. Phys.} \textbf{\bibinfo{volume}{28}},
  \bibinfo{pages}{258} (\bibinfo{year}{1958}).

\bibitem[{\citenamefont{Lifshitz and Slezov}(1959)}]{Lifshitz1959}
\bibinfo{author}{\bibfnamefont{I.~M.} \bibnamefont{Lifshitz}} \bibnamefont{and}
  \bibinfo{author}{\bibfnamefont{V.~V.} \bibnamefont{Slezov}},
  \bibinfo{journal}{Soviet Physics JETP} \textbf{\bibinfo{volume}{35}},
  \bibinfo{pages}{331} (\bibinfo{year}{1959}).

\bibitem[{\citenamefont{Slezov and Schmelzer}(1997)}]{Slezov1997}
\bibinfo{author}{\bibfnamefont{V.~V.} \bibnamefont{Slezov}} \bibnamefont{and}
  \bibinfo{author}{\bibfnamefont{J.}~\bibnamefont{Schmelzer}},
  \bibinfo{journal}{Phys.\ Solid State} \textbf{\bibinfo{volume}{39}},
  \bibinfo{pages}{1971} (\bibinfo{year}{1997}).

\bibitem[{\citenamefont{Mainville et~al.}(1997)\citenamefont{Mainville, Yang,
  Elder, and Sutton}}]{Mainville1997}
\bibinfo{author}{\bibfnamefont{J.}~\bibnamefont{Mainville}},
  \bibinfo{author}{\bibfnamefont{Y.~S.} \bibnamefont{Yang}},
  \bibinfo{author}{\bibfnamefont{K.~R.} \bibnamefont{Elder}}, \bibnamefont{and}
  \bibinfo{author}{\bibfnamefont{M.}~\bibnamefont{Sutton}},
  \bibinfo{journal}{Phys. Rev. Lett.} \textbf{\bibinfo{volume}{78}},
  \bibinfo{pages}{2787} (\bibinfo{year}{1997}).

\bibitem[{\citenamefont{Ng et~al.}(1996)\citenamefont{Ng, Lorber, Witz,
  Th\'{e}obald-Dietrich, Kern, and Gieg\'{e}}}]{Ng1996}
\bibinfo{author}{\bibfnamefont{J.~D.} \bibnamefont{Ng}},
  \bibinfo{author}{\bibfnamefont{B.}~\bibnamefont{Lorber}},
  \bibinfo{author}{\bibfnamefont{J.}~\bibnamefont{Witz}},
  \bibinfo{author}{\bibfnamefont{A.}~\bibnamefont{Th\'{e}obald-Dietrich}},
  \bibinfo{author}{\bibfnamefont{D.}~\bibnamefont{Kern}}, \bibnamefont{and}
  \bibinfo{author}{\bibfnamefont{R.}~\bibnamefont{Gieg\'{e}}},
  \bibinfo{journal}{J.\ Crystal Growth} \textbf{\bibinfo{volume}{168}},
  \bibinfo{pages}{50} (\bibinfo{year}{1996}).

\bibitem[{\citenamefont{Smith et~al.}(2003)\citenamefont{Smith, Kingsley,
  Maidanov, Rudavskii, Grigorev, Slezov, Poole, Saunders, and
  Cowan}}]{Smith2003}
\bibinfo{author}{\bibfnamefont{A.}~\bibnamefont{Smith}},
  \bibinfo{author}{\bibfnamefont{S.}~\bibnamefont{Kingsley}},
  \bibinfo{author}{\bibfnamefont{V.~A.} \bibnamefont{Maidanov}},
  \bibinfo{author}{\bibfnamefont{E.~Y.} \bibnamefont{Rudavskii}},
  \bibinfo{author}{\bibfnamefont{V.~N.} \bibnamefont{Grigorev}},
  \bibinfo{author}{\bibfnamefont{V.~V.} \bibnamefont{Slezov}},
  \bibinfo{author}{\bibfnamefont{M.}~\bibnamefont{Poole}},
  \bibinfo{author}{\bibfnamefont{J.}~\bibnamefont{Saunders}}, \bibnamefont{and}
  \bibinfo{author}{\bibfnamefont{B.}~\bibnamefont{Cowan}},
  \bibinfo{journal}{Phys.\ Rev.\ B} \textbf{\bibinfo{volume}{67}},
  \bibinfo{pages}{245314} (\bibinfo{year}{2003}).

\bibitem[{\citenamefont{Penzev et~al.}(2002)\citenamefont{Penzev, Ganshin,
  Grigor'ev, Maidanov, Rudavskii, Rybalko, Slezov, and Syrnikov}}]{Penzev2002}
\bibinfo{author}{\bibfnamefont{A.}~\bibnamefont{Penzev}},
  \bibinfo{author}{\bibfnamefont{A.}~\bibnamefont{Ganshin}},
  \bibinfo{author}{\bibfnamefont{V.}~\bibnamefont{Grigor'ev}},
  \bibinfo{author}{\bibfnamefont{V.}~\bibnamefont{Maidanov}},
  \bibinfo{author}{\bibfnamefont{E.}~\bibnamefont{Rudavskii}},
  \bibinfo{author}{\bibfnamefont{A.}~\bibnamefont{Rybalko}},
  \bibinfo{author}{\bibfnamefont{V.}~\bibnamefont{Slezov}}, \bibnamefont{and}
  \bibinfo{author}{\bibfnamefont{Y.}~\bibnamefont{Syrnikov}},
  \bibinfo{journal}{J.\ Low Temp.\ Phys.} \textbf{\bibinfo{volume}{126}},
  \bibinfo{pages}{151} (\bibinfo{year}{2002}).

\bibitem[{\citenamefont{Nagler et~al.}(1988)\citenamefont{Nagler,
  R.~F.~Shannon, Harkless, Singh, and Nicklow}}]{Nagler1988}
\bibinfo{author}{\bibfnamefont{S.~E.} \bibnamefont{Nagler}},
  \bibinfo{author}{\bibfnamefont{J.}~\bibnamefont{R.~F.~Shannon}},
  \bibinfo{author}{\bibfnamefont{C.~R.} \bibnamefont{Harkless}},
  \bibinfo{author}{\bibfnamefont{M.~A.} \bibnamefont{Singh}}, \bibnamefont{and}
  \bibinfo{author}{\bibfnamefont{R.~M.} \bibnamefont{Nicklow}},
  \bibinfo{journal}{Phys.\ Rev.\ Lett.} \textbf{\bibinfo{volume}{61}},
  \bibinfo{pages}{718} (\bibinfo{year}{1988}).

\bibitem[{\citenamefont{Greenberg et~al.}(1972)\citenamefont{Greenberg,
  Thomlinson, and Richardson}}]{Greenberg1972}
\bibinfo{author}{\bibfnamefont{A.~S.} \bibnamefont{Greenberg}},
  \bibinfo{author}{\bibfnamefont{W.~C.} \bibnamefont{Thomlinson}},
  \bibnamefont{and} \bibinfo{author}{\bibfnamefont{R.~C.}
  \bibnamefont{Richardson}}, \bibinfo{journal}{J.\ Low Temp.\ Phys.}
  \textbf{\bibinfo{volume}{8}}, \bibinfo{pages}{3} (\bibinfo{year}{1972}).

\bibitem[{\citenamefont{Cowan}(1997)}]{Cowan1997}
\bibinfo{author}{\bibfnamefont{B.}~\bibnamefont{Cowan}},
  \emph{\bibinfo{title}{Nuclear Magnetic Resonance and Relaxation}}
  (\bibinfo{publisher}{Cambridge University Press}, \bibinfo{year}{1997}).

\bibitem[{\citenamefont{Mullin}(1968)}]{Mullin1968}
\bibinfo{author}{\bibfnamefont{W.~J.} \bibnamefont{Mullin}},
  \bibinfo{journal}{Phys.\ Rev.\ Lett.} \textbf{\bibinfo{volume}{20}},
  \bibinfo{pages}{254} (\bibinfo{year}{1968}).

\bibitem[{\citenamefont{Edwards and Balibar}(1989)}]{Edwards1989}
\bibinfo{author}{\bibfnamefont{D.~O.} \bibnamefont{Edwards}} \bibnamefont{and}
  \bibinfo{author}{\bibfnamefont{S.}~\bibnamefont{Balibar}},
  \bibinfo{journal}{Phys.\ Rev.\ B} \textbf{\bibinfo{volume}{39}},
  \bibinfo{pages}{4083} (\bibinfo{year}{1989}).

\bibitem[{\citenamefont{Poole et~al.}(2006)\citenamefont{Poole, Saunders, and
  Cowan}}]{Poole2006}
\bibinfo{author}{\bibfnamefont{M.}~\bibnamefont{Poole}},
  \bibinfo{author}{\bibfnamefont{J.}~\bibnamefont{Saunders}}, \bibnamefont{and}
  \bibinfo{author}{\bibfnamefont{B.}~\bibnamefont{Cowan}},
  \bibinfo{journal}{Phys.\ Rev.\ Lett.} \textbf{\bibinfo{volume}{97}},
  \bibinfo{pages}{125301} (\bibinfo{year}{2006}).

\bibitem[{\citenamefont{Fraass and Simmons}(1987)}]{Fraass1987}
\bibinfo{author}{\bibfnamefont{B.~A.} \bibnamefont{Fraass}} \bibnamefont{and}
  \bibinfo{author}{\bibfnamefont{R.~O.} \bibnamefont{Simmons}},
  \bibinfo{journal}{Phys. Rev. B.} \textbf{\bibinfo{volume}{36}},
  \bibinfo{pages}{97} (\bibinfo{year}{1987}).

\bibitem[{\citenamefont{Grigor'ev}(1997)}]{Grigorev1997}
\bibinfo{author}{\bibfnamefont{V.~N.} \bibnamefont{Grigor'ev}},
  \bibinfo{journal}{Low Temp.\ Phys.} \textbf{\bibinfo{volume}{23}},
  \bibinfo{pages}{3} (\bibinfo{year}{1997}).

\bibitem[{\citenamefont{Ganshin et~al.}(1999)\citenamefont{Ganshin, Grigor'ev,
  Maidanov, Omelaenko, Penzev, Rudavskii, Rybalko, and Tokar}}]{Ganshin1999}
\bibinfo{author}{\bibfnamefont{A.~N.} \bibnamefont{Ganshin}},
  \bibinfo{author}{\bibfnamefont{V.~N.} \bibnamefont{Grigor'ev}},
  \bibinfo{author}{\bibfnamefont{V.~A.} \bibnamefont{Maidanov}},
  \bibinfo{author}{\bibfnamefont{N.~F.} \bibnamefont{Omelaenko}},
  \bibinfo{author}{\bibfnamefont{A.~A.} \bibnamefont{Penzev}},
  \bibinfo{author}{\bibfnamefont{E.}~\bibnamefont{Rudavskii}},
  \bibinfo{author}{\bibfnamefont{A.~S.} \bibnamefont{Rybalko}},
  \bibnamefont{and} \bibinfo{author}{\bibfnamefont{Y.~A.} \bibnamefont{Tokar}},
  \bibinfo{journal}{Fiz. Nizk. Temp.} \textbf{\bibinfo{volume}{25}},
  \bibinfo{pages}{796} (\bibinfo{year}{1999}).

\bibitem[{\citenamefont{Ohishi et~al.}(1998)\citenamefont{Ohishi, Yamamoto, and
  Suzuki}}]{Ohishi1998}
\bibinfo{author}{\bibfnamefont{K.}~\bibnamefont{Ohishi}},
  \bibinfo{author}{\bibfnamefont{H.}~\bibnamefont{Yamamoto}}, \bibnamefont{and}
  \bibinfo{author}{\bibfnamefont{M.}~\bibnamefont{Suzuki}},
  \bibinfo{journal}{J.\ Low Temp.\ Phys.} \textbf{\bibinfo{volume}{112}},
  \bibinfo{pages}{199} (\bibinfo{year}{1998}).

\bibitem[{\citenamefont{Jones}(2002)}]{Jones2002}
\bibinfo{author}{\bibfnamefont{R.~A.~L.} \bibnamefont{Jones}},
  \emph{\bibinfo{title}{Soft Condensed Matter}} (\bibinfo{publisher}{Oxford
  Master Series in Condensed Matter Physics}, \bibinfo{year}{2002}).

\bibitem[{\citenamefont{Koster et~al.}(1995)\citenamefont{Koster, Nagler,
  Adams, and Wignall}}]{Koster1995}
\bibinfo{author}{\bibfnamefont{J.~P.} \bibnamefont{Koster}},
  \bibinfo{author}{\bibfnamefont{S.~E.} \bibnamefont{Nagler}},
  \bibinfo{author}{\bibfnamefont{E.~D.} \bibnamefont{Adams}}, \bibnamefont{and}
  \bibinfo{author}{\bibfnamefont{G.~D.} \bibnamefont{Wignall}},
  \bibinfo{journal}{Mat. Res. Soc. Proc.} \textbf{\bibinfo{volume}{376}},
  \bibinfo{pages}{335} (\bibinfo{year}{1995}).

\end{thebibliography}

\end{document}